\documentclass[prl,twocolumn,showpacs]{revtex4}
\usepackage{graphicx}
\usepackage{amsmath}
\usepackage{amsfonts}
\usepackage{bm}
\usepackage{psfrag}

\renewcommand{\k}{{\mathbf{k}}}

\newcommand{\q}{{\mathbf{q}}}

\begin{document}

\title{Coherent Propagation of Polaritons in Semiconductor Heterostructures:\\ Nonlinear Pulse Transmission in Theory and Experiment}

\author{S. Schumacher}
\author{G. Czycholl}
\author{F. Jahnke}

\affiliation{Institute for Theoretical Physics,
             University of Bremen,
             28334 Bremen, Germany}

\author{I. Kudyk}
\author{L. Wischmeier}
\author{I. R\"uckmann}
\author{T. Voss}
\author{J. Gutowski}

\affiliation{Institute for Solid State Physics,
             Semiconductor Optics Group,
             University of Bremen,
             28334 Bremen, Germany}

\author{A. Gust}
\author{D. Hommel}

\affiliation{Institute for Solid State Physics,
             Semiconductor Epitaxy Group,
             University of Bremen,
             28334 Bremen, Germany}

\date{\today}


\begin{abstract}
The influence of coherent optical nonlinearities on polariton
propagation effects is studied within a theory-experiment comparison.
A novel approach that combines a microscopic treatment of the boundary
problem in a sample of finite thickness with excitonic and biexcitonic
nonlinearities is introduced.  Light-polarization dependent spectral
changes are analyzed for single-pulse transmission and pump-probe
excitation.
\end{abstract}

\maketitle

The propagation of light pulses which are coupled to the excitonic
resonances is a fundamental problem in semiconductor optics that has
been the subject of intense experimental and theoretical research.
The observations are dominated by the strong light-matter interaction
and its interplay with the inherent many-particle Coulomb interaction
of the electronic system. In the linear optical regime,
exciton polaritons give rise to effects like polariton beating in
the time-resolved pulse transmission \cite{Froehlich1991} or yield
strong modifications of excitonic transmission spectra, see e.g.
\cite{Tredicucci1993,Tignon2000,Schumacher2004}.  In the nonlinear
optical regime, incoherent saturation of the polariton resonances in
transmission spectra \cite{Neukirch1997,Betz2002} and in the amplitude
and phase of the time-resolved transmission \cite{Naegerl2001} as well
as transmission changes in pump-probe experiments \cite{Schaefer1997}
have been studied.

The above investigations are focused on samples where the thickness
corresponds to a few exciton Bohr radii where polariton effects are
most pronounced in the optical spectra.  In this case the interplay
of the induced excitonic polarization in the medium with the
propagating light field is strongly influenced by sample surfaces
which considerably complicates the theoretical description. In a
frequently used phenomenological approach \cite{Pekar1958} the
coupling of the exciton relative and center-of-mass (COM) motion at
the sample boundaries is neglected. Then the exciton COM motion is
subject to quantization effects in the confinement geometry while
the exciton relative motion is approximated by the result of the
infinitely extended medium. As a result of this approach, the
solution of the wave equation for the electromagnetic field requires
additional boundary conditions (ABCs) \cite{Pekar1958,Ting1975}
which are, however, not uniquely defined.  Unfortunately, in many
situations the choice of ABCs can influence the theoretical
predictions \cite{Schneider2001}. To avoid these ambiguities, the
use of microscopic boundary conditions within a non-local
semiconductor response has been discussed in
Refs.~\cite{Dandrea1982,Dandrea1990} and recently applied to the
linear optical regime \cite{Tignon2000,Muljarov2002,Schumacher2004}.
The description of optical nonlinearities combined with a
microscopic treatment of boundary conditions has been restricted so
far to the quantum-well (QW) limit where propagation effects lead to
radiative exciton broadening, which can be modified in radiatively
coupled QWs. Furthermore, excitonic nonlinearities in QWs have been
studied in connection with microcavity polaritons; for a review see
\cite{Khitrova1999}. In the opposite limit of the sample thickness
approaching the bulk limit nonlinear pulse propagation effects
\cite{Giessen1998,Foerstner2001} have been analyzed successfully
within a local approach for the semiconductor response.  The
decoupling of relative and COM exciton motion has also been used in
an earlier approach to study the influence of propagation effects on
four-wave mixing signals \cite{Schulze1995}.

In this letter we present a novel approach that describes the
combined influence of polariton propagation effects and coherent
excitonic and biexcitonic nonlinearities.  While the theoretical
understanding of the coherent nonlinear excitation dynamics in QW or
bulk systems is well developed, the additional influence of
propagation effects is missing especially in a regime where the
sample thickness exceeds the QW limit but the heterostructure
interfaces prevent bulk-like behavior. In this case typically
several well-resolved polariton modes contribute to the optical
properties and, as a result of our investigations, the Coulomb
interaction of the corresponding excitonic states modifies optical
nonlinearities.

Our approach is based on a direct solution of the two- and
four-particle Schr\"odinger equations for the exciton and biexciton
motion together with Maxwell's equations.  We apply microscopic
boundary conditions to avoid ambiguities due to ABCs in situations
where both material polarization as well as optical fields are
strongly influenced by the boundaries of the system.  A direct
comparison of nonlinear transmission spectra for a ZnSe/ZnSSe
heterostructure shows excellent agreement between theory and
experiment and supports the detailed analysis of the microscopic
model.  As another application, calculated pump-probe spectra for a GaAs/AlGaAs
heterostructure are presented.

In the coherent regime, excitonic and biexcitonic nonlinearities up to
third order in the optical field can be consistently described in
terms of the dynamics-controlled truncation (DCT) formalism
\cite{Axt1994a,Lindberg1994} which has been successfully applied in
the past to QW systems
\cite{Axt1995,Sieh1999,Schaefer2001,Kwong2001a,Donovan2001,Buck2004}.
We extend this formalism to layers with a finite thickness where the
sample boundaries still provide a confinement potential for the
electrons and holes. To account for the coupling of the exciton
relative and COM motion, independent coordinates for the electron and
hole motion in propagation direction are used while these carriers are
subject to Coulomb interaction (treated on the level of DCT) and to
the optical field.

We consider a two-band model with spin-degenerate conduction and
valence bands to describe the spectrally lowest interband transitions
for a semiconductor layer in a slab geometry \cite{Schneider2001}
with strain-split
light- and heavy-hole valence bands. The resonant contribution to
the spatially inhomogeneous polarization in the semiconductor,
\begin{align}\label{macpol}
\mathbf{P}(z,t)
={d}^\ast_{\text{e}\text{h}}\sum_{\k}\Big(p_{(\k,z,z)}^{-\frac{1}{2}-\frac{3}{2}}
\mathbf{e}_++p_{(\k,z,z)}^{+\frac{1}{2}+\frac{3}{2}}\mathbf{e}_-\Big),
\end{align}
is given in terms of the non-local excitonic transition amplitude
$p_{(\k,z_{\text{e}},z_{\text{h}})}^{\text{e}\text{h}}(t)=\big\langle
h_{\k}(z_{\text{h}})e_{\k}(z_{\text{e}})\big\rangle$ with the electron
$e_{\k}(z_{\text{e}})$ and hole $h_{\k}(z_{\text{h}})$ operators,
respectively.  According to the dipole selection rules, the
($-\frac{1}{2},-\frac{3}{2}$) and ($+\frac{1}{2},+\frac{3}{2}$)
transitions are independently driven by circular optical polarization
with unit vectors $\mathbf{e}_+$ and $\mathbf{e}_-$, respectively,
$\k$ is the in-plane carrier momentum (for the directions
possessing translational invariance) and $\text{e}$, $\text{h}$ are
quantum numbers that simultaneously denote the band index and the
$z$-component of the corresponding total angular momenta. A local
dipole-interaction
$\mathbf{d}_{\text{eh}}(\k,z_{\text{e}}-z_{\text{h}})=\mathbf{d}_{\text{eh}}\delta(z_{\text{e}}-z_{\text{h}})$
is used.  Applying the DCT scheme, an equation of motion for the
excitonic transition amplitude
$p_{(\k,z_{\text{e}},z_{\text{h}})}^{\text{e}\text{h}}$ is obtained
which is coupled to the biexcitonic correlation function,
\begin{align}\label{XXcorr}
b_{{\text{eh}}}^{{\text{e}}'{\text{h}}'} \,
^{(\k_2,z_2,\k_1,z_1)}_{(\k_4,z_4,\k_3,z_3)}&= \big\langle
h'_{\k_1}(z_1)e'_{\k_2}(z_2)h_{\k_3}(z_3)e_{\k_4}(z_4)\big\rangle
\nonumber
\\ -&\big\langle h'_{\k_1}(z_1)e'_{\k_2}(z_2)\big\rangle
\big\langle h_{\k_3}(z_3)e_{\k_4}(z_4)\big\rangle \nonumber
\\
+&\big\langle h'_{\k_1}(z_1)e_{\k_4}(z_4)\big\rangle \big\langle
h_{\k_3}(z_3)e'_{\k_2}(z_2)\big\rangle,
\end{align}
that itself obeys a four-particle Schr\"odinger equation with coherent
source terms \cite{Schaefer2002}.  The resulting closed set of
equations contains all mean-field (Hartree-Fock) as well as
biexcitonic contributions to the semiconductor response up to third
order in the optical field. The evaluation of these material equations
is performed in the exciton eigenbasis. As an important ingredient of
our approach, exciton eigenfunctions
$\phi_m(\k,z_{\text{e}},z_{\text{h}})$ are determined {\em for the
given confinement geometry} to include the additional non-local space
dependence. In contrast to Ref.~\cite{Muljarov2002} we use
eigenfunctions that individually fulfill the boundary conditions of
the system \cite{Schumacher2002,Schumacher2004}. Then every
eigenfunction corresponds to a polariton resonance in the linear
optical spectrum and a truncation of the expansion can be justified
with the considered spectral range while the boundary conditions of
the inhomogeneous system remain fully satisfied. The excitonic
transition amplitude and the singlet ($\lambda=-1$) and triplet
($\lambda=+1$) contributions to the biexcitonic correlation function
are expressed as
\begin{align}\label{Xexpansion}
p^{\text{eh}}_{(\k,z_{\text{e}},z_{\text{h}})}(t)=\sum_m{p_m^{\text{eh}}(t)\phi_m(\k,z_{\text{e}},z_{\text{h}})},
\end{align}
\begin{align}\label{XXexpansion}
b_{{\text{eh}}}^{{\text{e}}'{\text{h}}'\lambda}&\,^{(\k_2,z_{2},\k_1,z_{1})}_{(\k_4,z_{4},\k_3,z_{3})}(t)
= \nonumber \\ \nonumber
\sum_{nm}\Big[&\phi_n(\alpha\k_4+\beta\k_3,z_4,z_3)\phi_m(\alpha\k_2+\beta\k_1,z_2,z_1)
\\ \nonumber &\times b_{nm}^{{\text{ehe}}'{\text{h}}'\lambda}(\k_4-\k_3,t)
\\ \nonumber
-\lambda&\phi_n(\alpha\k_2+\beta\k_3,z_2,z_3)\phi_m(\alpha\k_4+\beta\k_1,z_4,z_1)
\\
&\times
b_{nm}^{{\text{ehe}}'{\text{h}}'\lambda}(\k_2-\k_3,t)\Big]
\end{align}
where $p_m^{\text{eh}}(t)$ and $b^{ehe'h'\lambda}_{nm}(\q,t)$ are the
time-dependent expansion coefficients.
$\alpha=m^\ast_{\text{h}}/M^\ast$ and $\beta=m^\ast_{\text{e}}/M^\ast$
are the ratio of the hole and the electron masses to the total exciton
mass $M^\ast=m^\ast_{\text{e}}+m^\ast_{\text{h}}$, respectively.  This
treatment generalizes the eigenfunction expansion in the QW limit
discussed in \cite{Takayama2002}.  For a self-consistent description
of the light propagation in a semiconductor material, Maxwell's equations
are solved directly \cite{Schneider2001} together with the excitonic
and the biexcitonic dynamics.

\begin{figure}
\includegraphics[scale=1.03]{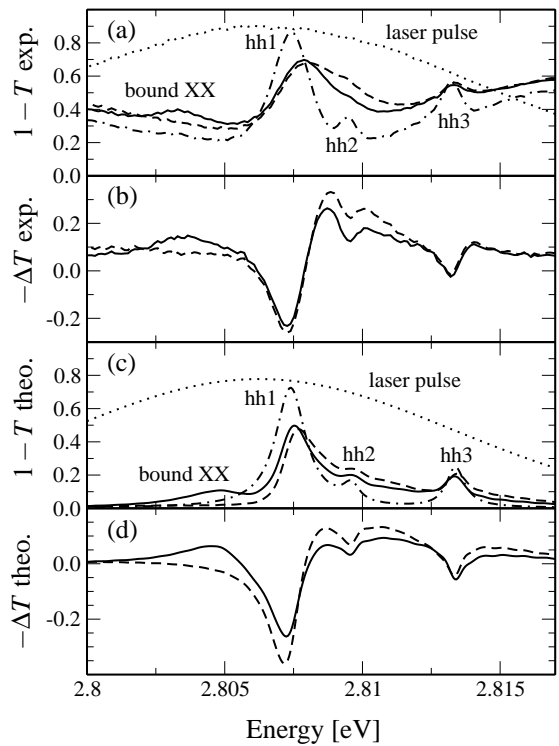}
\caption{(a) Dashed-dotted line: Linear transmission $T$ (depicted
as $1-T$) for a pulse energy of $1.2\,\mathrm{pJ}$ for the
$20\,\mathrm{nm}$ ZnSe sample. Dotted line: Spectral shape of the
$110\,\mathrm{fs}$ laser pulse. Nonlinear transmission for a pulse
energy of $12.2\,\mathrm{pJ}$ for linear (solid line) and circular
(dashed line) light polarization. (b) Optically induced transmission
changes $\Delta T$ for linear (solid line) and circular (dashed
line) light polarization, corresponding to the data in (a).  (c),
(d) Theoretical results corresponding to (a) and (b), without the
influence of Fabry-Perot effects (see discussion in the text).}
\label{Fig2}
\end{figure}

In Fig.~\ref{Fig2} the theoretical and experimental results are
shown for the transmission of a single light pulse through a
high-quality $20\,\mathrm{nm}$ ZnSe layer sandwiched between two
ZnSSe cladding layers. The sample was pseudomorphically grown by
molecular-beam epitaxy on a (001)GaAs substrate. Further details on
growth and sample characterization as well as on modeling of the
finite-height confinement potentials at the ZnSe/ZnSSe interfaces
are described in \cite{Schumacher2004}. The opaque GaAs substrate
was carefully removed by grinding and subsequent chemical etching to
permit measurements in transmission geometry. The sample was mounted
onto a glass plate and kept on a cold finger in a microcryostat at a
temperature of 4 K. A frequency-doubled self-mode-locked Ti:sapphire
laser with $110\,\mathrm{fs}$ (FWHM) pulse duration and
$83\,\mathrm{MHz}$ repetition rate was used as excitation source for
the spectrally resolved but time integrated transmission
experiments.  The maximum of the spectral pulse profile was set
close to the spectral position of the first heavy-hole (hh1)
polariton mode to excite the whole spectral region from the
exciton-biexciton transition up to the hh2 and hh3 polariton
resonances as shown by the dotted line in Fig.~\ref{Fig2}a.  Linear
or circular light polarization was selected using a Pockels cell.
The transmitted signal was analyzed with a spectrometer and was
recorded with a CCD camera.

The measured linear transmission spectrum in the vicinity of the
heavy-hole (hh) exciton resonance, depicted as dashed-dotted line in
Fig.~\ref{Fig2}a, exhibits three pronounced polariton resonances
denoted by hh1-hh3. With the microscopic treatment of boundary
conditions, these polariton modes are perfectly reproduced in
Fig.~\ref{Fig2}c (without adjustable parameters like a dead-layer
thickness, the sample thickness has been independently determined
using x-ray diffraction \cite{Schumacher2004}).  Results for
nonlinear transmission spectra are shown in Figs.~\ref{Fig2}a
(experiment) and c (theory) where linear and circular light
polarization has been used. For better comparison, the optically
induced transmission changes $\Delta
T=T_{\text{nonlinear}}-T_{\text{linear}}$ are given in
Figs.~\ref{Fig2}b and d, for the experiment and theory,
respectively. The outermost surfaces between cladding layer and air
lead to a decreased transmission as well as to weak Fabry-Perot
effects in the experiment \cite{Schumacher2004}. In order to avoid
additional parameters, the theory only includes intrinsic polariton
effects. The constant offset in the measured nonlinear transmission
results from slight intensity drifts of the applied laser pulses.
These effects are of minor importance for the discussion here, since
they just cause a nearly constant offset in the investigated
transmission changes. The appearance of a bound biexciton (XX)
resonance for linear optical polarization, which is absent for
circular optical polarization according to its spin-singlet
symmetry, is clearly identified in the theory-experiment comparison.
The spectral changes at the polariton resonances result from
mean-field (Hartree-Fock) as well as from biexcitonic correlations.
For circular light polarization only the biexcitonic continuum
states with electronic triplet configuration are excited whereas for
linear light polarization both singlet and triplet biexcitonic
states contribute to the transmitted signal. In particular, an
advantage of the theoretical approach applied here is the
simultaneous inclusion of the bound biexciton state and the
exciton-exciton scattering continuum. The latter one is essential to
reproduce the broad background on the high-energy side of the
polariton resonances in the nonlinear transmission spectra. As a
minor aspect, the biexciton binding energy is slightly
underestimated by the truncation of the exciton basis, similar to
the result in \cite{Buck2004} for a QW system.

\begin{figure}
\includegraphics[scale=1.03]{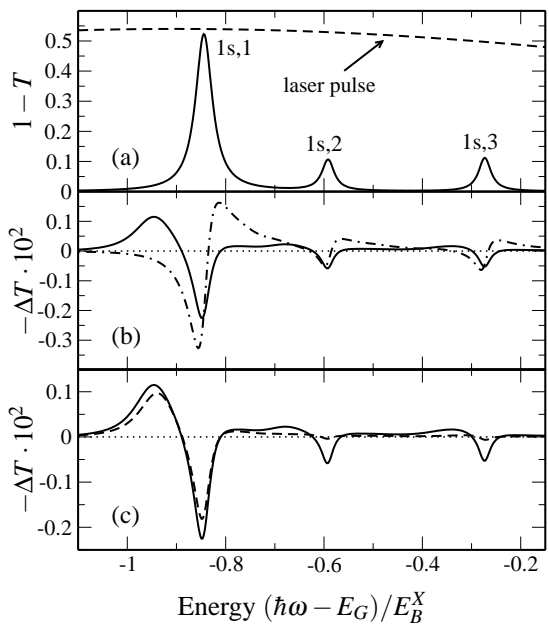}
\caption{(a) Solid line: Calculated linear transmission spectrum $T$
(depicted as $1-T$) for a GaAs layer with 5 exciton Bohr radii
thickness. Dashed line: Spectral shape of a $120\,\mathrm{fs}$ laser
pulse. (b) Differential probe transmission $\Delta T$ for opposite
circular (solid line), and co-circular (dashed-dotted line)
polarization of pump and probe pulses.  (c) Differential probe
transmission $\Delta T$ for opposite circular polarization including
all Coulomb terms (solid line) and without Coulomb interaction of
different polaritons (dashed line).}\label{Fig3}
\end{figure}

To provide further insight into the nature of coherent nonlinear
polariton saturation, we also studied the pump-probe excitation for
a GaAs layer embedded between AlGaAs barriers \footnote{Material
parameters are: $m^\ast_{\text{e}}=0.067\,\mathrm{m_0}$,
$m^\ast_{\text{h}}=0.457\,\mathrm{m_0}$ for effective electron and
hole masses, with the bare electron mass $m_0$, the background
refractive index $n_{\text{bg}}=3.71$, a dephasing constant
$\gamma=0.06\,\mathrm{meV}$, the bulk GaAs band-gap energy
$E_G=1.42\,\mathrm{eV}$, and a dipole coupling constant
$d_{\text{eh}}/e_0=5\,\mathrm{\AA}$.}. This part of the paper is
intended to demonstrate the potential of our theory within another
application and to stimulate further experiments for this material
system. In a good approximation, the carriers in the GaAs layer are
confined by infinitely high potential barriers in the $z$-direction
\cite{Schneider2001}. The solid line in Fig.~\ref{Fig3}a shows the
linear optical transmission $T$ through a GaAs layer of $5$ exciton
Bohr radii thickness.  The energy $\hbar\omega$ is given relative to
the bulk band-gap energy $E_G$ and normalized to the corresponding
exciton binding energy $E_B^X$.  For the chosen layer thickness, the
confinement of carriers in the $z$-direction yields three polariton
resonances of the 1s exciton in the displayed part of the spectrum
which are labeled by consecutive numbers.

For the pump-probe excitation, two $120\,\mathrm{fs}$ laser pulses
(from slightly different directions) without time-delay are
considered. The dashed line in Fig.~\ref{Fig3}a corresponds to the
pulse spectrum.  Figure~\ref{Fig3}b shows changes in the probe-pulse
transmission that are induced by the pump pulse for opposite circular
(solid line) and co-circular (dashed-dotted line) polarization of pump
and probe pulses. The probe pulse enters the excitonic polarization in
linear order only. For the pump pulse a Rabi energy
$d_{\text{eh}}|\mathbf{E}_{\text{pump}}|=0.01\,\mathrm{E_B^X}$ is used
which ensures a consistent description within our $\chi^{(3)}$
theory. The transmission changes around the higher (1s,2 and 1s,3)
polariton re\-sonances in Fig.~\ref{Fig3}b are similar to those around
the lowest one (1s,1) but with a decreased amplitude. A similar
dependence on the light polarization has been reported for the
differential probe absorption around the 1s exciton resonance in a QW
system \cite{Sieh1999}. For opposite-circular polarization, the
pump-induced changes in the probe transmission are exclusively
determined by biexcitonic correlations \cite{Sieh1999}; no mean-field
effects contribute. This configuration is chosen for the analysis of
the Coulomb interaction between polaritons in states with different
spatial distribution.  The dashed line in Fig.~\ref{Fig3}c shows the
result where Coulomb interaction that couples different eigenfunctions
in the two-exciton product basis of Eq.~(\ref{XXexpansion}) is
artificially switched off. We encounter only a slight change of the
nonlinearities around the lowest polariton resonance (1s,1) whereas
for higher peaks (1s,2 and 1s,3) the influence of the pump pulse
almost vanishes.  Therefore, Coulomb interaction between exciton
states with different spatial distribution (corresponding to different
polariton resonances) turns out to be the main
source for transmission changes of higher polariton states.

In summary, a theory for the nonlinear polariton dynamics in
semiconductor heterostructures has been presented which consistently
includes (i) propagation effects with microscopic boundary conditions
for the induced material polarization as well as for the optical
fields and (ii) excitonic and biexcitonic coherent nonlinearities
previously studied only in QWs or one-dimensional model systems.  The
application to nonlinear transmission spectra for a ZnSe/ZnSSe
heterostructure shows excellent agreement between theory and
experiment. The influence of light-polarization dependent excitonic
and biexcitonic nonlinearities has also been demonstrated for a
pump-probe excitation of a GaAs heterostructure.  Coulomb interaction
of polariton states with different spatial distribution turns out to
strongly influence the nonlinear optical transmission.

\begin{acknowledgments}

The authors acknowledge a grant for CPU time from the John von
Neumann Institute for Computing at the Forschungszentrum J\"ulich.
We thank G. Alexe for the sample characterization via x-ray
diffraction.

\end{acknowledgments}

\bibliography{./literature_short}
\bibliographystyle{apsrev}

\end{document}